\documentclass[11pt,a4paper]{article}
\usepackage{amsmath} 
\usepackage{amssymb} 
\usepackage{bm}  
\usepackage{color}  

\def\be{\begin{equation}}
\def\ee{\end{equation}}
\def\ba{\begin{eqnarray}}
\def\ea{\end{eqnarray}}
\def\onehalf{{\textstyle{\frac{1}{2}}}}

\begin{document}

\begin{center}
{\Large \bf de Sitter transitivity, conformal}
\vskip 0.1cm
{\Large \bf transformations and conservation laws}
\vskip 0.5cm
{\bf J. G. Pereira, A. C. Sampson and L. L. Savi}\\
{\it Instituto de F\'{\i}sica Te\'orica, Universidade Estadual Paulista \\
Caixa Postal 70532-2, 01156-970 S\~ao Paulo, Brazil}
\end{center}


\vskip 0.5cm
\begin{quote}
{\bf Abstract.}~{\footnotesize Minkowski spacetime is transitive under ordinary translations, a transformation that do not have matrix representations. The de Sitter spacetime, on the other hand, is transitive under a combination of translations and proper conformal transformations, which do have a matrix representation. Such matrix, however, is not by itself a de Sitter generator: it gives rise to a conformal re-scaling of the metric, a transformation not belonging to the de Sitter group, and in general not associated with diffeomorphisms in spacetime. When dealing with variational principles and Noether's theorem in de Sitter spacetime, therefore, it turns out necessary to regularise the transformations in order to eliminate the {conformal re-scaling of the metric}.

}
\end{quote}

\section{Introduction}
\label{intro}

Spacetimes with constant sectional curvature are maximally symmetric in the sense that they can lodge the highest possible number of Killing vectors \cite{weinberg}. Minkowski spacetime $M$, with metric $\eta_{\mu \nu}$ and vanishing curvature, is the simplest one. Its kinematic group is the Poin\-ca\-r\'e group ${\mathcal P} = {\mathcal L} \oslash {\mathcal T}$, the semi-direct product of Lorentz ${\mathcal L}$ and the translation group ${\mathcal T}$. It is a homogeneous space defined by the quotient
\[
M = {\mathcal P}/{\mathcal L}.
\]
The invariance of $M$ under the transformations of ${\mathcal P}$ reflects its uniformity. The Lorentz subgroup provides an isotropy around a given point of $M$, and the translation symmetry enforces this isotropy around any other point. This is the meaning of homogeneity: all points of $M$ are ultimately equivalent under translations.

In addition to Minkowski, there are two other maximally symmetric four-dimensional spacetimes \cite{ellis}. One is de Sitter, with topology $R^1 \times S^3$ and (let us say) positive sectional curvature. The other is anti-de Sitter, with topology $S^1 \times R^3$ and negative sectional curvature. As hyperbolic spaces both have negative Gaussian curvature. Here we will be interested in the de Sitter spacetime $dS(4,1)$, whose kinematic is ruled by the de Sitter group $SO(4,1)$. It is defined by the quotient
\[
dS(4,1) = SO(4,1) / {\mathcal L}.
\]
Together with Minkowski, de Sitter can be considered a fundamental spacetime, a stage where physics is to be developed \cite{evolve}. Of course, in order to be physically relevant it must be solution to Einstein equation. However, it is more fundamental than Einstein equation in the sense that, as a quotient space, it is known {\it a priori}, independently of general relativity. Like Minkowski, it is a homogeneous spacetime. The notion of homogeneity, however, which enforces the Lorentz symmetry in all other points of spacetime, is completely different: instead of translations, any two points of this spacetime are connected by a combination of translation and proper conformal transformation \cite{dSsr0}. The purpose of this paper is to explore the consequences of this difference for the notion of motion in de Sitter spacetime, as well as for the ensuing conservation laws.

\section{Lorentz transformations revisited}
\label{sec:LorenCase}

To begin with, and for the sake of comparison, let us consider the well-known case of the Lorentz transformations. The generators of infinitesimal Lorentz transformations are made up of two parts: a derivative (or orbital) part, which is the same for all fields, and a matrix (or spin) part, which depends on the spin of the field under consideration. The explicit form of these two generators, denoted respectively by $L_{\alpha \beta}$ and $S_{\alpha \beta}$, can be obtained by computing the Lie derivative of the field along the Killing vectors of the Lorentz transformations. As an illustration, let us consider the case of a vector field $\psi_\mu$. A Lorentz transformation in Minkowski spacetime can be written in the form
\be
\delta_{L} x^\gamma = \onehalf \varepsilon^{\alpha \beta} \, \xi^{~\gamma}_{(\alpha \beta)},
\label{LoreTrans}
\ee
where $\varepsilon^{\alpha \beta} = - \varepsilon^{\beta\alpha}$ are the {\em constant} parameters and
\be
\xi^{~\gamma}_{(\alpha \beta)} = 
\eta_{\alpha \delta} x^\delta \delta^\gamma_\beta -
\eta_{\beta \delta} x^\delta \delta^\gamma_\alpha
\label{LoreKiVec}
\ee
are the Killing vectors of the Lorentz group. The change of a vector field $\psi_\mu$ under such transformation is given by the Lie derivative of $\psi_\mu$ along the direction of the Killing vectors~$\xi^{~\gamma}_{(\alpha \beta)}$:
\be
\delta^0_{L} \psi_\mu \equiv \psi'(x) - \psi(x) = (\mathcal{L}_{L} \psi)_\mu.
\ee
Computing the Lie derivative, we get
\be
\delta^0_{L} \psi_\mu = -
\onehalf \varepsilon^{\alpha\beta} \xi^{~\gamma}_{(\alpha \beta)} \partial_\gamma \psi_\mu -
\onehalf \varepsilon^{\alpha\beta} \partial_\mu \xi^{~\gamma}_{(\alpha \beta)} \psi_\gamma.
\label{dSTrans0}
\ee
This transformation can be rewritten in the form
\be
\delta^0_{L} \psi_\mu = -
\onehalf \varepsilon^{\alpha\beta} L_{\alpha \beta} \psi_\mu -
\onehalf \varepsilon^{\alpha\beta} (S_{\alpha \beta})_\mu{}^\gamma \psi_\gamma,
\label{dSTrans0again}
\ee
where
\be
L_{\alpha \beta} \equiv \xi^{~\gamma}_{(\alpha \beta)} \partial_\gamma =
\eta_{\alpha\delta} x^\delta \partial_\beta - \eta_{\beta\delta} x^\delta \partial_\alpha
\label{lab}
\ee
are the orbital generators and
\be
({S}_{\alpha\beta})_\mu{}^\gamma \equiv
\partial_\mu \xi^{~\gamma}_{(\alpha \beta)} =
\eta_{\alpha\mu} \delta_\beta^\gamma - \eta_{\beta\mu} \delta_\alpha^\gamma
\label{dSM1}
\ee
stand for the spin-1 matrix representation of the Lorentz generators \cite{ramond5}.
Transformation (\ref{dSTrans0}) can be schematically rewritten as
\be
\delta^0_{L} \psi_\mu = -
\onehalf \varepsilon^{\alpha\beta} ( L_{\alpha \beta} +
S_{\alpha \beta} ) \, \psi_\mu.
\label{dSTrans0bis}
\ee
The combination of generators
\be
J_{\alpha \beta} = L_{\alpha \beta} + S_{\alpha \beta}
\label{completeG}
\ee
satisfies the commutation relation
\be
\left[J_{\alpha \beta}, J_{\gamma \delta}\right] = 
\eta_{\beta\gamma} J_{\alpha\delta} + \eta_{\alpha\delta} J_{\beta\gamma} -
\eta_{\beta\delta} J_{\alpha\gamma} - \eta_{\alpha\gamma} J_{\beta\delta},
\label{LorenAlge}
\ee
with $L_{\alpha \beta}$ and $S_{\alpha \beta}$ satisfying, each one, the same algebra and commuting with each other:
\be
\left[S_{\alpha \beta}, L_{\gamma \delta}\right] = 0.
\ee

The field transformation coming from the change of the {\em argument} (the so-called transport term) is generated by the orbital representation
\be
\delta^a_{L} \psi_\mu \equiv \psi_\mu(x') - \psi_\mu(x) = \onehalf \varepsilon^{\alpha\beta} L_{\alpha \beta} \psi_\mu.
\ee 
The {\em total} change in the field, therefore, is generated by the matrix representation
\be
\psi'_\mu(x') - \psi_\mu(x) = - \onehalf \varepsilon^{\alpha\beta} (S_{\alpha \beta})_\mu{}^\gamma \psi_\gamma.
\ee
This is the transformation appearing in special relativity, which says how a vector field is seen from two different observers attached to frames $K'$ and $K$. 

\section{Minkowski transitivity: ordinary translations}

Before considering the transitivity of the de Sitter spacetime, it is instructive to recall the case of Minkowski spacetime, which is well-known to be transitive under ordinary translations. A global translation in this spacetime is written as
\be
\delta_{P} x^\gamma = \delta^{\;\gamma}_{(\alpha)} \, \varepsilon^\alpha
\label{OrTrKilVec}
\ee
where $\delta^{\;\gamma}_{(\alpha)}$ are the translational Killing vectors and $\varepsilon^\alpha$ are the constant transformation parameters. The corresponding generators of infinitesimal transformations are
\be
P_\alpha = \delta^{\;\gamma}_{(\alpha)} \partial_\gamma.
\label{P}
\ee
The behavior of a vector field $\psi_\mu$ under such transformations is given by the Lie derivative of $\psi_\mu$ along the direction of the Killing vectors $\delta^{\;\gamma}_{(\alpha)}$:
\be
\delta^0_{P} \psi_\mu \equiv (\mathcal{L}_P \psi)_\mu = -
\varepsilon^\alpha \delta^{\;\gamma}_{(\alpha)} \partial_\gamma \psi_\mu.
\label{LieDerVec0}
\ee
From Eq.~(\ref{dSM1}) we see that the spin matrix generators of the Lorentz transformations come from the dependence of the Killing vectors on the spacetime coordinates. The reason why ordinary translations do not have a matrix representation is that the translation Killing vectors $\delta^{~\gamma}_{(\alpha)}$ are constant, and consequently the corresponding matrix representation vanishes:
\[
({\Sigma}_{\alpha})_\mu{}^\gamma
\equiv \partial_{\mu} \delta^{~\gamma}_{(\alpha)} = 0.
\] 
On the other hand, the field transformation coming from the change of the spacetime point (argument), is
\be
\delta_{P}^a \psi_\mu \equiv \psi_\mu(x') - \psi_\mu(x) = 
\varepsilon^\alpha \delta^{\;\gamma}_{(\alpha)} \partial_\gamma \psi_\mu.
\label{TransInvari5a}
\ee
The total change in the field, therefore, vanishes identically:
\be
\psi'_\mu(x') - \psi_\mu(x) \equiv \delta_P^0 \psi_\mu + \delta_P^a \psi_\mu = 0.
\label{TransInvari5}
\ee
This is an expected result because, since Minkowski is transitive under spacetime translations, a global translation corresponds to a mere redefinition of the spacetime origin, which of course does not affect local fields \cite{livro}.

\section{Transitivity of de Sitter spacetime}

Differently from Minkowski, the de Sitter spacetime is transitive under a combination of translation and proper conformal transformation, the so-called de Sitter ``translation''. In this section we explore further some consequences of this property.

\subsection{Generators of de Sitter ``translations''}

In terms of the stereographic coordinates $\{x^\mu\}$, and considering a parameterisation appropriate for small values of $\Lambda$ \cite{gursey}, a de Sitter ``translation'' is written as
\be
\delta_\Pi x^\gamma = \xi^{\;\gamma}_{(\alpha)} \, \varepsilon^\alpha
\label{dStransCoord5}
\ee
where $\varepsilon^\alpha$ are the constant transformation parameters and
\be
\xi^{\;\gamma}_{(\alpha)} =
\delta^\gamma_\alpha -
\frac{1}{4l^2}(2 \eta_{\alpha \delta} x^\delta  x^\gamma -
\sigma^2 \delta^\gamma_\alpha)
\label{pia}
\ee
are the de Sitter ``translation'' Killing vectors, with 
$\sigma^2$ the Lorentz-invariant quadratic form $\sigma^2 = \eta_{\mu \nu} \, x^\mu x^\nu$ and $l$ the de Sitter length-parameter (or pseudo-radius).
The corresponding generators of infinitesimal transformations are \cite{dSsr0}
\be
\Pi_\alpha = \xi^{\;\gamma}_{(\alpha)} \partial_\gamma \equiv
\partial_\alpha - \frac{1}{4l^2} (2 \eta_{\alpha \delta} x^\delta  x^\gamma -
\sigma^2 \delta^\gamma_\alpha) \partial_\gamma
\label{pia2}.
\ee
They satisfy the commutation relation \cite{zimer}
\be
\left[\Pi_{\alpha}, \Pi_{\beta}\right] = l^{-2} \, L_{\alpha \beta}
\ee
with $L_{\alpha \beta}$ the orbital Lorentz generators~(\ref{lab}). This shows that the de Sitter ``translations'' are not really translations, but rotations (hence the quotation marks).


The behavior of a vector field $\psi_\mu$ under a de Sitter ``translation'' is given by the Lie derivative of $\psi_\mu$ along the direction of the Killing vectors~$\xi^{\;\gamma}_{(\alpha)}$:
\be
\delta^0_\Pi \psi_\mu \equiv (\mathcal{L}_\Pi \psi)_\mu = -
\varepsilon^\alpha \, \xi^{\;\gamma}_{(\alpha)} \partial_\gamma \psi_\mu -
\varepsilon^\alpha \partial_\mu \xi^{\;\gamma}_{(\alpha)} \psi_\gamma.
\ee
It can be rewritten in the form
\be
\delta^0_\Pi \psi_\mu = -
\varepsilon^\alpha \Pi_{\alpha} \psi_\mu -
\varepsilon^\alpha ({\Sigma}_{\alpha})_\mu{}^\gamma \psi_\gamma.
\label{dStransVecS}
\ee
The first term on the right-hand side represents the action of the derivative generators. In analogy to the Lorentz case discussed in Section~\ref{sec:LorenCase}, the second term should be interpreted as the action of the {\em matrix representation of the proper conformal generators}, whose explicit form is \cite{dStransRep}
\be
({\Sigma}_{\alpha})_\mu{}^\gamma
\equiv \partial_{\mu} \xi^{\;\gamma}_{(\alpha)} =
\frac{1}{2l^2}( \eta_{\mu\beta} x^\beta \delta_\alpha^\gamma -
\eta_{\alpha\mu} x^\gamma - \eta_{\alpha \beta} x^\beta \delta_\mu^\gamma).
\label{dSM2S}
\ee

However, there is a problem with this interpretation: the matrices ${\Sigma}_{\alpha}$ do not satisfy the de Sitter algebra. In fact, as a simple computation shows, 
\be
[\Sigma_\alpha, \Sigma_\beta] = l^{-2} X_{\alpha \beta},
\label{sigsig}
\ee
where $X_{\alpha \beta}$ represent spurious terms in relation to the de Sitter algebra. {\em This means that ${\Sigma}_{\alpha}$ alone is not a generator of de Sitter transformations}. In spite of this fact, (\ref{dStransVecS}) is a de Sitter transformation. To see that, let us rewrite it in the schematic form
\be
\delta^0_\Pi \psi_\mu = -
\varepsilon^\alpha \Delta_{\alpha} \psi_\mu,
\label{dStransVecSb}
\ee
where
\be
\Delta_{\alpha} = \Pi_\alpha + \Sigma_\alpha.
\label{dSGenVec}
\ee
It is then easy to verify that $\Delta_{\alpha}$ does satisfy the de Sitter algebra, that is,
\be
[\Delta_\alpha, J_{\beta \gamma} ] = \eta_{\alpha \beta} \Delta_\gamma -
\eta_{\alpha \gamma} \Delta_\beta
\label{TotaldS Al2}
\ee
and
\be
[\Delta_\alpha, \Delta_\beta ] = l^{-2} \, J_{\alpha \beta}.
\label{TotaldS Al3}
\ee
We see from these relations that, even though $\Sigma_\alpha$ is not, $\Delta_{\alpha}$ is a de Sitter generator. It is interesting to observe that in the decomposition
\be
[\Delta_\alpha, \Delta_\beta] = [\Pi_\alpha, \Pi_\beta] + [\Pi_\alpha, \Sigma_\beta] +
[\Sigma_\alpha, \Pi_\beta] + [\Sigma_\alpha, \Sigma_\beta],
\label{DelDel}
\ee 
because the derivative generator $\Pi_\alpha$ and the matrix generator $\Sigma_\alpha$ do not commute,
\be
[\Pi_\alpha, \Sigma_\beta] + [\Sigma_\alpha, \Pi_\beta] = l^{-2} \left(S_{\alpha \beta} -
X_{\alpha \beta} \right),
\ee
the second and third terms on the right-hand side of (\ref{DelDel}) give rise to spurious terms that exactly compensate the spurious terms coming from the last commutator, given by Eq.~(\ref{sigsig}), yielding the de Sitter commutation relation (\ref{TotaldS Al3}). It is also interesting to observe that, when $\Sigma_\alpha$ is included in the ``translational'' de Sitter generator, as in Eq.~(\ref{dSGenVec}), the de Sitter algebra turns out to be written, not with the orbital Lorentz generators $L_{\alpha \beta}$, but with the complete generators $J_{\alpha \beta} = L_{\alpha \beta} + S_{\alpha \beta}$. This shows that, although not itself a de Sitter generator, the matrix $\Sigma_\alpha$ has a relevant role for the algebraic structure of the de Sitter group.

\subsection{Transformations generated by $\Sigma_\alpha$}

Let us explore in more details the transformations generated by $\Sigma_\alpha$. Under the de Sitter ``translation'' (\ref{dStransCoord5}), the transformation of a vector field coming from the change of the spacetime point (or argument) is
\be
\delta ^a_\Pi \psi_\mu \equiv \psi_\mu(x') - \psi_\mu(x) = 
\varepsilon^\alpha \Pi_{\alpha} \psi_\mu.
\label{ImportantTransS}
\ee
This means that, similarly to the Lorentz group, the total transformation is found to be generated by the matrix generators
\be
\psi'_\mu(x') - \psi_\mu(x) \equiv 
\delta^0_\Pi \psi_\mu + \delta ^a_\Pi \psi_\mu = - \varepsilon^\alpha (\Sigma_\alpha)_\mu{}^\gamma \psi_\gamma.
\label{TotaldStrans}
\ee
There is a crucial difference, though: since $\Sigma_\alpha$ alone is not a de Sitter generator, this is not a de Sitter transformation.

Let us consider now the case of the metric tensor $g_{\mu \nu}$, whose transformation is given by its Lie derivative along the de Sitter Killing vectors $\xi^{\;\gamma}_{(\alpha)}$
\be
\delta^0_\Pi g_{\mu \nu} = -
\varepsilon^\alpha \Pi_\alpha g_{\mu \nu} -
\varepsilon^\alpha (\Sigma_\alpha)_\mu{}^\gamma g_{\gamma \nu} -
\varepsilon^\alpha (\Sigma_\alpha)_\nu{}^\gamma g_{\mu \gamma}
\label{dSTMetricTransS}
\ee
where we have already used the definition (\ref{dSM2S}). Of course, since $\xi^{\;\gamma}_{(\alpha)}$ is a Killing vector, this transformation vanishes
\be
\delta^0_\Pi g_{\mu \nu} = 0.
\ee
On the other hand, the transformation of the metric tensor due to the change in the argument is given by
\be
\delta^a_\Pi g_{\mu \nu} \equiv g_{\mu \nu}(x') - g_{\mu \nu}(x) =
\varepsilon^\alpha \Pi_\alpha g_{\mu \nu}(x).
\ee
The total transformation of the metric is then found to be
\be
g'_{\mu \nu}(x') - g_{\mu \nu}(x) = -
\varepsilon^\alpha (\Sigma_\alpha)_\mu{}^\gamma g_{\gamma \nu} -
\varepsilon^\alpha (\Sigma_\alpha)_\nu{}^\gamma g_{\mu \gamma}.
\label{TotaldStransMet}
\ee
Substituting $\Sigma_\alpha$ as given by Eq.~(\ref{dSM2S}), it assumes the form
\be
g'_{\mu \nu}(x') = \omega^2 g_{\mu \nu}(x)
\label{TotaldStransMet2}
\ee
with
\be
\omega^2 = 1 + \frac{\varepsilon_\alpha x^\alpha}{l^2}.
\label{ConforFacS5}
\ee
For the metric tensor, therefore, the transformation generated by $\Sigma_\alpha$ is just an infinitesimal conformal re-scaling of the metric, with $\omega^2$ as conformal factor. 

\subsection{de Sitter ``translation'' and transitivity}

The study of conformal geometry, that is, of the set of all metrics obtained through a conformal transformation from the physical metric $g_{\mu \nu}$, is equivalent to the study of causal relationships in spacetime (see Ref.~\cite{ellis}, page 180). This equivalence is related to the fact that the light-cone, which defines the causal structure of spacetime, is invariant under conformal re-scalings of the metric. Such transformations, however, are not associated with diffeomorphisms in spacetime (see Ref.~\cite{wald5}, page 445). In fact, the invariance of a physical system under a metric conformal re-scaling is not related to any conservation law through Noether theorem. {Maxwell theory, for example, is invariant under a conformal re-scaling of the metric, but no conserved current exists associated to this invariance.\footnote{Notice that this is different from invariance under proper conformal transformations, which leads to the conservation of the proper conformal current \cite{coleman5}.}}

On the other hand, considering that the de Sitter spacetime is transitive under de Sitter ``translations'', a global de Sitter ``translation'' should represent a mere redefinition of the origin of spacetime, and consequently it should not affect local fields. In particular, the metric should remain invariant
\be
g_{\mu \nu}'(x') - g_{\mu \nu}(x) = 0,
\ee
instead of transforming according to (\ref{TotaldStransMet2}). This means that, in what concerns the transitivity of the de Sitter spacetime, {\em one must consider a de Sitter ``translation'' up to a conformal re-scaling of the field}. For a vector field, such transformation can be written in the form
\be
\delta^0_\Pi \psi_\mu = (\mathcal{L}_\Pi \psi)_\mu + 
\varepsilon^\alpha (\Sigma_\alpha)_\mu{}^\gamma\psi_\gamma.
\ee
The addition of the last term removes the undesired part of the transformation---that is, the part that generates a conformal re-scaling of the field---yielding a {\em genuine} de Sitter ``translation'', that is, a transformation defining the transitivity of the de Sitter spacetime. Using Eq.~(\ref{dStransVecS}), such transformation is found to be
\be
\delta^0_\Pi \psi_\mu = -
\varepsilon^\alpha \Pi_{\alpha} \psi_\mu.
\label{LieDerVecB}
\ee
Taking into account that the transformation of a vector field coming from the change of the spacetime point is
\be
\delta ^a_\Pi \psi_\mu = 
\varepsilon^\alpha \Pi_{\alpha} \psi_\mu,
\label{ImportantTransSbis}
\ee
the total transformation is now given by
\be
\psi'_\mu(x') - \psi_\mu(x) \equiv 
\delta^0_\Pi \psi_\mu + \delta ^a_\Pi \psi_\mu = 0,
\ee
as appropriate for a transformation defining the transitivity of a spacetime.

For the metric tensor, the genuine de Sitter ``translation'' has the form
\be
\delta^0_\Pi g_{\mu \nu} = (\mathcal{L}_\Pi g)_{\mu \nu} + 
\varepsilon^\alpha (\Sigma_\alpha)_\mu{}^\gamma g_{\gamma \nu} +
\varepsilon^\alpha (\Sigma_\alpha)_\nu{}^\gamma g_{\mu \gamma}.
\label{RegMetrTrans}
\ee
Using (\ref{dSTMetricTransS}), it reduces to
\be
\delta^0_\Pi g_{\mu \nu} = -
\varepsilon^\alpha \Pi_\alpha g_{\mu \nu}.
\label{dSTMetricTransSb}
\ee
On the other hand, considering that the transformation in the metric tensor due to the change of the argument is
\be
\delta^a_\Pi g_{\mu \nu} =
\varepsilon^\alpha \Pi_\alpha g_{\mu \nu}(x),
\label{RegdSTrans}
\ee
the total transformation is
\be
g'_{\mu \nu}(x') - g_{\mu \nu}(x) = 0,
\ee
a result consistent with the transitivity properties of the de Sitter spacetime. The genuine de Sitter ``translations'', that is, the de Sitter ``translations'' up to a conformal re-scaling of the metric, are the relevant transformations to be used whenever obtaining conservation laws from Noether's theorem, as well as field equations from a variational principle.
 
\section{An example: de Sitter kinematics} 

Let us consider a particle of mass $m$, whose action functional is given by
\be
S = -\, m c \int_a^b ds
\label{graviaction}
\ee
where $ds = (g_{\alpha \beta} \, dx^\alpha dx^\beta)^{1/2}$, with $g_{\alpha \beta}$ the de Sitter metric. Under the spacetime variation (\ref{dStransCoord5}), which takes into account the transitivity properties of the de Sitter spacetime, the action transforms according to
\begin{equation}
\delta S = - mc \int_{b}^{a} \left[ \onehalf \, \delta_{\Pi} (g_{\alpha\gamma}) \, u^{\alpha} dx^{\gamma} +
g_{\alpha\beta} \, u^{\alpha} \, \delta_{\Pi} (dx^{\beta}) \right],
\label{eq:2}
\end{equation}
with $u^\alpha = d x^\alpha/ds$ the particle four-velocity. Substituting $\delta_{\Pi} (g_{\alpha\gamma})$ as given by Eq.~(\ref{RegdSTrans}) and using the identity $\delta_{\Pi} (dx^{\beta}) = d (\delta_{\Pi} x^{\beta})$ in the last term, we get
\begin{equation}
\delta S = - mc \int_{b}^{a} \left[ \onehalf \, \xi^{\beta}_{\rho} \partial_\mu \, g_{\alpha\gamma} \, u^{\alpha} dx^{\gamma} \, \varepsilon^{\rho} +
g_{\alpha\beta} \, u^{\alpha} \, d (\delta_{\Pi} x^{\beta}) \right].
\label{eq:2bis}
\end{equation}
Integrating the last term by parts and neglecting the surface term, the variation assumes the form
\begin{equation}
\delta S = - mc \int_{b}^{a} \left[ \frac{1}{2} \frac{\partial g_{\alpha\gamma}}{\partial x^{\beta}} \, u^{\alpha} u^{\gamma} -
\frac{d}{ds} \left( g_{\alpha\beta} \, u^{\alpha} \right)  \right]
\xi^{\beta}_{\rho} \, \varepsilon^{\rho} ds.
\label{eq:3}
\end{equation}
After some algebraic manipulation, it reduces to
\be
\delta S = m c \int_{b}^{a} \left[ u^{\gamma}
\nabla_{\gamma} u^{\beta} \, \xi^{\rho}_{\beta} \right] \varepsilon_{\rho} \, ds
\ee
with $\nabla_\gamma$ a covariant derivative in the Christoffel connection of the metric $g_{\alpha\beta}$. Defining the anholonomic four-velocity
\be
U^\rho = \xi^{\rho}_{\beta} \, u^{\beta}
\ee
it can be rewritten in the form
\be
\delta S = m c \int_{b}^{a} \left[ u^{\gamma}
\nabla_{\gamma} U^{\rho} -
\textstyle{\frac{1}{2}} u^\beta u^\gamma (\nabla_\beta \xi^{\rho}_{\gamma} +
\nabla_\gamma \xi^{\rho}_{\beta} ) \right] \varepsilon_{\rho} \, ds.
\label{62}
\ee
Using the definition~(\ref{dSM2S}), it becomes
\be
\delta S = m c \int_{b}^{a} \Big[ u^{\gamma}
\nabla_{\gamma} U^{\rho} -
\textstyle{\frac{1}{2}} u^\beta u^\gamma \big[ (\Sigma_\gamma)_\beta{}^\rho +
(\Sigma_\beta)_\gamma{}^\rho \big] \Big] \varepsilon_{\rho} \, ds.
\label{63}
\ee

Now, as discussed in the previous section, the terms involving the matrix operators $\Sigma_\beta$ represent a conformal resealing of the metric in the action variation. In fact, as a direct computation shows\footnote{It is interesting to note that Eq.~(\ref{ConforKE}) is similar to what is usually called the {\em conformal Killing equation}. See, for example, Ref.~\cite{wald5}, page 444.}
\be
(\Sigma_\gamma)_\beta{}^\rho + (\Sigma_\beta)_\gamma{}^\rho \equiv
\nabla_\beta \xi^{\rho}_{\gamma} +
\nabla_\gamma \xi^{\rho}_{\beta} = 
{\textstyle{\frac{1}{2}}} (\nabla^\nu \xi^\rho_\nu ) g_{\beta \gamma} =
-  \frac{x^\rho}{l^2} \, g_{\beta \gamma}.
\label{ConforKE}
\ee
Substituting into Eq.~(\ref{63}), we get
\be
\delta S = m c \int_{b}^{a} \Big[ u^{\gamma}
\nabla_{\gamma} U^{\rho} +
\frac{x^\rho}{2l^2} \Big] \varepsilon_{\rho} \, ds.
\label{64}
\ee
The second term in the integrand represents just an infinitesimal conformal re-scaling of the metric
\be
ds \to \omega \, ds
\ee
with
\be
\omega^2 = 1 + \frac{x^\rho \varepsilon_{\rho}}{l^2},
\ee
which is exactly the conformal factor~(\ref{ConforFacS5}). Up to a conformal re-scaling of the metric, therefore, the action variation is
\be
\delta S = m c \int_{b}^{a} u^{\gamma}
\nabla_{\gamma} U^{\rho} \varepsilon_{\rho} ds.
\label{67}
\ee
Taking into account the arbitrariness of the parameter $\varepsilon_{\rho}$, the invariance of the action yields
\be
\frac{d U^\rho}{ds} +
\Gamma^\rho{}_{\mu \gamma} \, U^\mu \, u^\gamma = 0. 
\label{GendSGeod}
\ee
This equation represents the particle trajectories in de Sitter spacetime. Owing to the fact that these trajectories are consistent with the transitivity properties of the de Sitter spacetime, any two points of this space will be connected by a trajectory of this family.\footnote{It is important to mention that the usual geodesics of the de Sitter metric are unable to connect all points of the spacetime; see, for example, Ref.~\cite{ellis}, page 126.} For this reason, they can be considered the true ``geodesics'' of the de Sitter spacetime \cite{dSgeode}. Furthermore, like in ordinary special relativity, they coincide with the four-momentum conservation. In fact, they can be written in the form
\be
\frac{d \pi^\rho}{ds} +
\Gamma^\rho{}_{\mu \gamma} \, \pi^\mu \, u^\gamma = 0, 
\label{GendSGeodBis}
\ee
where
\be
\pi^\rho = \xi^{\rho}_{\beta} \, p^{\beta}
\ee
is the de Sitter four-momentum of the particle, with $p^{\beta} = m c u^\beta$ the ordinary four-momentum.

\section{Final remarks}

The spacetime transitivity is intimately related to the notion of motion. In Minkowski, for example, any two points are connected by ordinary translation. This means that motion in such spacetime is described by trajectories whose points are connected to each other by a spacetime translation. On the other hand, the de Sitter spacetime is transitive under a combination of translation and proper conformal transformation. In this case, therefore, motion is described by trajectories whose points are connected to each other by a combination of translation and proper conformal transformation---the so-called de Sitter ``translation". 

Differently from ordinary translations, which do not have matrix representations, the de Sitter ``translations'' do have matrix representations. In fact, analogously to the Lorentz case, the generators of infinitesimal de Sitter ``translations''
\be
\Delta_\alpha = \Pi_\alpha + \Sigma_\alpha
\ee
are made up of two parts: a derivative part $\Pi_\alpha$, which has the same form for all fields, and a matrix part $\Sigma_\alpha$, whose explicit form depends on the spin of the field under consideration. However, there is a fundamental difference in relation to the Lorentz case: $\Pi_\alpha$ and $\Sigma_\alpha$ do not commute, and $\Sigma_\alpha$ alone does not satisfy the de Sitter algebra. Namely, it is not a de Sitter generator. In the specific case of the metric tensor, $\Sigma_\alpha$ is found to generate a conformal rescaling of the metric, which is not a transformation belonging to the de Sitter group.

Such transformation, however, is not associated with diffeomorphisms in spacetime. Notice, for example, that there is no a conserved quantity related to the invariance of a given system under a conformal rescaling of the metric. The transformation defining the transitivity of the de Sitter spacetime, therefore, must be regularised in order to eliminate the conformal rescaling of the metric. The resulting transformation can be considered a genuine de Sitter ``translation'' in the sense that it is the transformation that defines the transitivity of the de Sitter spacetime. It is the transformation to be used whenever dealing with Noether's theorem or variational principle in de Sitter spacetime --- as well as in locally de Sitter spacetimes \cite{wise}. 

\section*{Acknowledgements}
The authors would like to thank FAPESP, CAPES and CNPq for partial financial support.


\end{document}